\documentstyle[12pt]{article}
\newcommand{\nn}{\nonumber}
\newcommand{\bd}{\begin{document}}
\newcommand{\ed}{\end{document}}
\newcommand{\bc}{\begin{center}}
\newcommand{\ec}{\end{center}}
\newcommand{\be}{\begin{eqnarray}}
\newcommand{\ee}{\end{eqnarray}}
\renewcommand{\thefootnote}{\alph{footnote}}
\newcommand{\se}{\section}
\newcommand{\sse}{\subsection}
\newcommand{\bi}{\bibitem}
\def\figcap{\section*{Figure Captions\markboth
     {FIGURECAPTIONS}{FIGURECAPTIONS}}\list
     {Figure \arabic{enumi}:\hfill}{\settowidth\labelwidth{Figure 999:}
     \leftmargin\labelwidth
     \advance\leftmargin\labelsep\usecounter{enumi}}}
\let\endfigcap\endlist \relax
\input psfig

\begin{document}

\begin{titlepage}

 \vskip 0.5in
 \null
\begin{center}
 \vspace{.15in}
{\LARGE {\bf The
Muon Anomalous Magnetic Moment
from a Generic Charged Higgs with SUSY
}}\\
\vspace{1.0cm}  \par
 \vskip 2.1em
 {\large
  \begin{tabular}[t]{c}
{\bf Chuan-Hung Chen$^a$ and C.~Q.~Geng$^b$}
\\
\\
   {\sl ${}^a$Department of Physics, National Cheng Kung University}
\\   {\sl  $\ $Tainan, Taiwan,  Republic of China }
\\
\\
{\sl ${}^b$Department of Physics, National Tsing Hua University}
\\  {\sl  $\ $ Hsinchu, Taiwan, Republic of China }
\\
   \end{tabular}}
 \par \vskip 5.3em

\date{\today}
 {\Large\bf Abstract}
\end{center}

 We study the contribution of a generic charged Higgs $(H^+)$
to the muon anomalous magnetic moment $a_{\mu}$
with the SUSY soft breaking parameters. We find out that the deviation
 between the experimental data and the predicted SM value on $a_{\mu}$
 can be explained by the two-loop charged Higgs diagrams
even with $m_{H^+}\sim 400\ GeV$.

\end{titlepage}

It is believed that the muon anomalous magnetic moment, $a_\mu\equiv
(g_\mu-2)/2$, would provide precision tests of the standard model (SM) and
probe for new physics \cite{Czarnecki:2001pv}. Recently, it has been
measured at BNL \cite{Brown:2001mg} with the data
\begin{eqnarray}  \label{exp}
a_\mu^{exp}&=&116~ 592~ 023 (151)\times 10^{-11}~.
\end{eqnarray}
The experimental value in Eq. (\ref{exp}) differs from that in the SM \cite
{Brown:2001mg} despite of the several different theoretical predictions from
hadronic contributions \cite{hd1,hd2}. In Ref. \cite{Brown:2001mg}, it was
reported that
\begin{eqnarray}
\Delta a_\mu\equiv a_\mu^{{\rm exp}}-a_\mu^{{\rm SM}}=426\pm 165 \times
10^{-11}\,,  \label{dev}
\end{eqnarray}
while a recent calculation \cite{hd2} based on a different estimation from
the hadronic part gave
\begin{eqnarray}
\Delta a_\mu=375\pm170\times 10^{-11}\,.  \label{dev'}
\end{eqnarray}
The values in Eqs. (\ref{dev}) and (\ref{dev'}) indicate a window for new
physics at $2.6\sigma$ and $2.2\sigma$ levels, which are translated into
\begin{eqnarray}
215 \times 10^{-11} \leq \Delta a_\mu \leq 637 \times 10^{-11} \ (90\%\ {\rm %
CL})\,,  \label{newphys}
\end{eqnarray}
and
\begin{eqnarray}
159 \times 10^{-11} \leq \Delta a_\mu \leq 599 \times 10^{-11} \ (90\%\ {\rm %
CL})\,,  \label{newphys'}
\end{eqnarray}
respectively. It is clear that both ranges in Eqs. (\ref{newphys}) and (\ref
{newphys'}) suggest the existence of new physics beyond the SM. However, one
must caution about this less than $3\sigma$ result until the experiment of
E821 at BNL is completed, which should increase the statistical significance
at more than $6\sigma$ level \cite{Hughes}, and the theoretical
uncertainties from the hadronic part in $a_\mu^{{\rm SM}}$ are further
reduced.

Recently, various models, such as those with SUSY, scalar bosons, and extra
dimensions, which could lead to $\Delta a_\mu=O(400\times 10^{-11})$ have
been explored \cite{Czarnecki:2001pv,models,Haber,Chang}. In particular, it
is discussed extensively to use scalar Higgs bosons in SUSY-like theories as
the viable candidates. In Refs. \cite{Haber,Chang}, the possibilities of
using light neutral Higgs bosons with a large $tan\beta$ to account $\Delta
a_\mu$ at the one- and two-loop levels were studied. It is known that a
large $tan\beta$ is interesting theoretically since the unification of
bottom and tau Yukawa couplings and the explanation of the top to the bottom
mass ratio are realized in GUTs if $tan\beta\sim O(50)$ \cite{GUTs}. With
this possible large $tan\beta$, it is found that the mass for the scalar
boson has to be less than $5\ GeV$ \cite{Haber} and that for the
pseudoscalar $75\ GeV$ with including the Barr-Zee type \cite{BZ} of the
two-loop diagrams \cite{Chang}. One may conclude that $\Delta a_\mu$ cannot
arise from either a scalar or pseudoscalar in the minimal supersymmetric
model (MSSM) due to the experimental limits on the scalar and pseudoscalar
masses, which are in the ranges of $85-95\ GeV$ \cite{spml,PDG,hunter}.

In this paper, we would like to examine whether it is possible to use a
charged Higgs boson in SUSY models to induce $\Delta a_\mu$ beyond the one
loop level. It is known that the challenge with a charged Higgs in theories
is how to escape the constraint from the experimental value of $%
B(B\rightarrow X_{s}\gamma)=2.85\pm 0.41\times 10^{-4}$ \cite{CLEOBCP4}
which is consistent with that of $3.29\pm0.33\times 10^{-4}$ predicted in
the SM. Following the analysis of \cite{CDGG,CHml} with the next-to-leading
order (NLO) QCD corrections, the lower limit on the charged Higgs mass in
the two-Higgs doublet model (model II) is $450\ GeV$. However, the bound
can be
released in the framework of supersymmetric theories because of the somewhat
cancellation between the particle and its superpartner \cite{BG}. In Refs.
\cite{DGG} and \cite{CGNW}, it has been demonstrated that the NLO
contributions in a SUSY model with a sufficient large $tan\beta$ may be
comparable as that from the leading order (LO). Thus, with choosing a proper
sign of the Higgs mass mixing parameter $\mu$, the charged Higgs and
chargino contributions to $B\rightarrow X_s \gamma$ are suppressed.
Furthermore, SUSY models without R-parity, such as those by including $\mu_i
L_i H_d$, called bilinear terms, in the superpotential, can also allow a
charged Higgs as light as $80$ GeV by requiring the chargino mass $%
m_{\chi^{\pm}}> 90 GeV$ \cite{DTV}. Another scenario to evade the constraint
is proposed in Ref. \cite{MR}, in which the b-quark mass is induced from
radiative corrections so that the coupling $H^{+}\bar{t}_L b_R$ is generated
from higher order effects and thus, the Wilson coefficient of $C_7$ for $%
b\rightarrow s\gamma$ is suppressed by $1/tan^2\beta$. This also implies
that the bound on the charged Higgs mass can be lower without a fine tuning.
Hence, the light charged Higgs is still viable in some of SUSY models and it
could be reachable in future collider searches. In the following we shall
concentrate on a generic charged Higgs in models with SUSY, whose lower mass
limit is only constrained by the LEP experiments \cite{LEPchm}, $i.e.$,
\begin{eqnarray}
m_{H^+} &>& 80.5\ GeV\,.  \label{Chml}
\end{eqnarray}

The one-loop charged Higgs contribution to $a_\mu$ was studied previously
\cite{lautrup,ch1} and it was found that to accommodate the value of $\Delta
a_\mu$ in Eq. (\ref{newphys}) or (\ref{newphys'}), $m_{H^{+}}$ has to be
less than a few $GeV$ even with a large $\tan\beta$ \cite{Haber}. This
contribution is clearly negligible if one uses the limit in Eq. (\ref{Chml}%
). For the contribution to $\Delta a_\mu$ from the Barr-Zee type of the
two-loop diagrams \cite{BZ} with the charged Higgs in the loops similar to
the one in Ref. \cite{2edm}, we find that it is still small.

Since the one and two loop diagrams mentioned above involve only the well
known transition elements in which all couplings are almost fixed except $%
\tan \beta$, it seems to be impossible to generate a sizable $\Delta a_{\mu}
$ via loops with a charged Higgs. However, it is interesting to ask whether
there would exist some enhancement factors in some SUSY models with the
charged Higgs so that $\Delta a_{\mu}$ could be large. In fact, as we are
going to show next, such possibility could be realized by considering
two-loop diagrams in which the charged Higgs couples to squarks but not
quarks, and with introducing the SUSY soft breaking parameters for the
effects of the SUSY broken in the low energy.

\begin{figure}[tbp]
\hspace{2.5cm} \psfig{figure=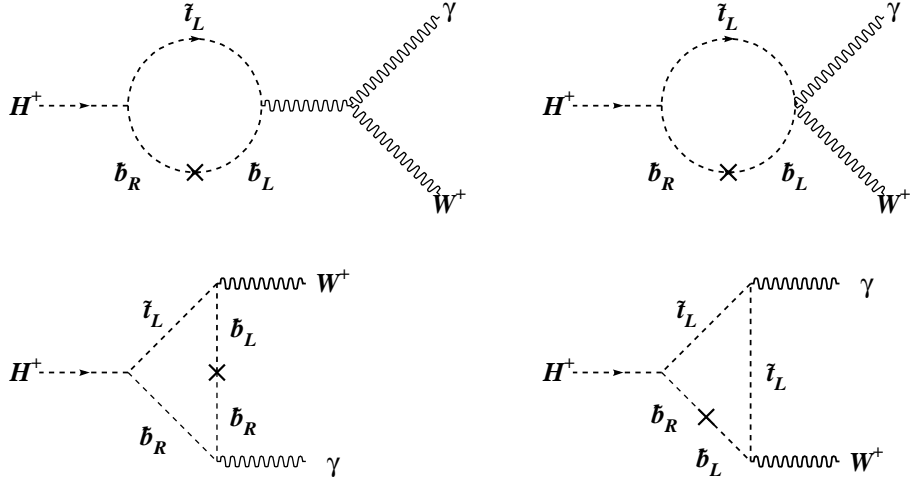,height=2.5in }
\caption{ Feynman diagrams for effective vertices of $H^{+}-\gamma-W^{+}$
where squarks are in the internal loops. }
\label{hwg}
\end{figure}

We start with the relevant couplings of the charged Higgs to squarks, the
trilinear soft breaking terms, given by
\begin{eqnarray}
{\cal {L}}_{soft}&=& A_{U}Y^{U}_{ij} \tilde{Q}_{i}H_{U}\tilde{U}^c
+A_{D}Y^{D}_{ij} \tilde{Q}_{i}H_{D}\tilde{D}^c\,,  \label{soft}
\end{eqnarray}
where $Y^{U,D}_{ij}$ denote the Yukawa couplings with i and j being the
flavor indices, and $A_{U,D}$ are the SUSY soft breaking parameters. From
Eq. (\ref{soft}), we see that the coupling for $H^+\tilde{t}^{*}_L \tilde{b}%
_R$ is $\sim m_bA_btan\beta$. With the terms in Eq. (\ref{soft}), the
effective vertex of $H^{+}-\gamma-W^+$ can be induced as shown in Figure \ref
{hwg} in which squarks are in the loops.

In SUSY models, the main contributions to $\Delta a_\mu$ are with $%
chargino-sneutrino$ and $neutralino-slepton$ couplings in loops \cite{SUSY}.
One can show that the former will become dominant, which is proportional to $%
(m_{\mu}/m_{SUSY})^2tan\beta$ if $tan\beta $ is large \cite{SUSY} where $%
m_{SUSY}$ denotes the the mass of the sneutrino or chargino. For $%
tan\beta>>1 $, $\Delta a_\mu $ can set a lower bound on the mass of the
sneutrino while the chargino is light. To concentrate on the charged Higgs
effect, we assume that both sneutrino and slepton masses are large enough so
that their effects are negligible for $\Delta a_\mu$, but we still need a
light chargino and squarks to satisfy the constraint from $b\rightarrow
s\gamma$.

The effective Lagrangian which describes the interaction of the charged
Higgs to squarks and leptons in terms of their weak eigenstates is
\[
{\cal L}_{H^{+}}=\frac{g}{\sqrt{2}m_{W}}\left[ m_{t}\mu \tilde{Q}_{ti}^{*}%
\tilde{Y}_{ij}\tilde{Q}_{bj}+m_{l}\tan \beta \bar{\nu}_{l}P_{R}\ell \right]
H^{+}+h.c.
\]
with
\begin{equation}
\tilde{Y}=\left[
\begin{tabular}{ll}
$-\hat{m}_{W}^{2}\sin 2\beta +\hat{m}_{b}^{\prime }\hat{m}_{b}\tan \beta +%
\hat{m}_{t}\cot \beta $ & $\hat{m}_{b}^{\prime }\left( 1-\hat{A}_{b}\tan
\beta \right) $ \\
\multicolumn{1}{c}{$1-\hat{A}_{t}\cot \beta $} & $2\hat{m}_{b}/\sin 2\beta $%
\end{tabular}
\right]   \label{susymatrix}
\end{equation}
where $\tilde{Q}_{t}^{*}=(\tilde{t}_{L}^{*},\tilde{t}_{R}^{*})$ and $\tilde{Q%
}_{b}^{T}=(\tilde{b}_{L},\tilde{b}_{R})$ are the stop and sbottom, the
parameters with a hat are renormalized by the $\mu $ parameter except $\hat{m%
}_{W}^{2}=m_{W}^{2}/m_{t}\mu $ and $\hat{m}_{b}^{\prime }=m_{b}/m_{t},$ and $%
\hat{A}_{t,b}$ are related to the SUSY soft breaking terms, respectively. 
 From Eq. (\ref{susymatrix}), we see that $\tilde{Y}_{LR}\sim
\hat{A}_b\tan\beta$, which contains not only $\tan \beta $ but also a
large factor from the soft SUSY breaking parameters. The relevent squark
mass matrix can be expressed by \cite{ER}
\be
M_{\tilde{q}}^{2}=\left[
\begin{tabular}{ll}
$m_{\tilde{q}_{L}}^{2}$ & $m_{q}\mu \hat{a}_{q}$ \\
$m_{q}\mu \hat{a}_{q}$ & $m_{\tilde{q}_{R}}^{2}$%
\end{tabular}
\right]
\ee
with
\be
m_{\tilde{q}_{L}}^{2} &=&M_{\tilde{q}_{L}}^{2}+m_{q}^{2}+m_{Z}^{2}\cos
2\beta \left( I_{q}^{3}-Q_{q}\sin ^{2}\theta _{W}\right) , 
\nn\\
m_{\tilde{q}_{R}}^{2} &=&M_{\tilde{q}_{R}}^{2}+m_{q}^{2}+Q_{q}m_{Z}^{2}\cos
2\beta \sin ^{2}\theta _{W}, 
\nn\\
\hat{a}_{b} &=&\hat{A}_{b}-\tan \beta , 
\nn\\
\hat{a}_{t} &=&\hat{A}_{t}-\cot \beta
\ee
where $M_{\tilde{q}_{L,R}}^{2}$ arise from the SUSY broken effects. Hence,
the physical sbottom and stop masses 
can be found as
\begin{equation}
m_{\tilde{q}_{1,2}}^{2}=\frac{1}{2}\left( m_{\tilde{q}_{L}}^{2}+m_{\tilde{q}%
_{R}}^{2}\mp \sqrt{\left( m_{\tilde{q}_{L}}^{2}-m_{\tilde{q}_{R}}^{2}\right)
^{2}+4m_{q}^{2}\mu ^{2}\hat{a}_{q}^{2}}\right) .  \label{sm}
\end{equation}
In our discussion, we take 
$m_{\tilde{q}_{L}}^{2}\simeq m_{\tilde{q}_{R}}^{2}
\simeq m_{\tilde{q}}^{2}$ so that 
$m_{\tilde{q}_{1,2}}^{2}\simeq m_{\tilde{q}}^{2}\mp \left| m_{q}\mu
\hat{a}_{q}\right|$.
For ensuring the squark masses  being positive, we require that
$\left|m_{q}\mu \hat{a}_{q}\right| <m_{\tilde{q}}^{2}$.
By using  $\mu\simeq 2\ TeV$, $\hat{A}_{q}\simeq -2$ and $\tan \beta
\simeq 40$, and choosing  $m_{\tilde{b}}\simeq 645\ GeV$
and $m_{\tilde{t}}\simeq 820\ GeV$, one can show that
the lightest sbottom and stop can be $\sim 100\ GeV$.

 From Figure 1, the gauge invariant form of the effective coupling for $%
H^{+}-\gamma -W^{+}$ with squarks in the loops is expressed as \cite{PHI}
\begin{equation}
\Gamma ^{\mu \nu }\left( q^{2}\right) =N_{c}\frac{c_{\tilde{t}}s_{\tilde{b}%
}\alpha _{em}em_{b}\mu \hat{A}_{b}^{*}}{4m_{W}\pi \sin ^{2}\theta _{W}}%
\int_{0}^{1}dx\frac{x\left( 1-x\right) \left( Q_{t}\left( 1-x\right)
+Q_{b}x\right) }{\left( m_{\tilde{t}_{1}}^{2}-q^{2}x\right) \left(
1-x\right) +m_{\tilde{b}_{1}}^{2}x}\left[ q^{\mu }k^{\nu }-q\cdot kg^{\mu
\nu }\right]  \label{gamma}
\end{equation}
where $N_{c}=3$, $Q_{t}=2/3$ and $Q_{b}=-1/3$ are stop and sbottom charges,
and $c_{\tilde{t}}=\cos \theta _{\tilde{t}}$ and $s_{\tilde{b}}=\sin \theta
_{\tilde{b}}$ express the mixings of left and right squarks in $\tilde{t}%
_{1}=c_{\tilde{t}}\tilde{t}_{L}+s_{\tilde{t}}\tilde{t}_{R}$ and $\tilde{b}%
_{1}=c_{\tilde{b}}\tilde{b}_{L}+s_{\tilde{b}}\tilde{b}_{R}$, respectively.
In the following discussions, we shall assume that $\tilde{t}_{1}$ and $%
\tilde{b}_{1}$ are the lightest squarks and use $c_{\tilde{t}}s_{\tilde{b}%
}=0.5$ \cite{PDG}. From Eq. (\ref{gamma}), the two-loop contributions to $%
\Delta a_{\mu }$ are found to be
\begin{eqnarray}
\triangle a_{l} &=&\frac{N_{c}c_{\tilde{t}}s_{\tilde{b}}\alpha _{em}^{2}\tan
^{2}\beta }{16\pi ^{2}\sin ^{4}\theta _{W}}\frac{m_{b}\mu Re(\hat{A}_{b}^{*})%
}{m_{W}^{2}}\frac{m_{l}^{2}}{m_{H^{+}}^{2}}  \nonumber \\
&&\times \left[ Q_{t}J\left( \frac{m_{W}^{2}}{m_{H^{+}}^{2}},\frac{m_{\tilde{%
t}_{1}}^{2}}{m_{H^{+}}^{2}},\frac{m_{\tilde{b}_{1}}^{2}}{m_{H^{+}}^{2}}%
\right) +Q_{b}J\left( \frac{m_{W}^{2}}{m_{H^{+}}^{2}},\frac{m_{\tilde{b}%
_{1}}^{2}}{m_{H^{+}}^{2}},\frac{m_{\tilde{t}_{1}}^{2}}{m_{H^{+}}^{2}}\right)
\right]  \label{g-2}
\end{eqnarray}
with
\[
J\left( a,b,c\right) =\frac{1}{1-a}\left( I(b,c)-I(\frac{b}{a},\frac{c}{a}%
)\right)
\]
where
\[
I(b,c)=\int_{0}^{1}dx\frac{x(1-x)^{2}}{(b-x)(1-x)+cx}\ln \frac{x(1-x)}{%
b(1-x)+cx}.
\]

It is worth to mention that by replacing the incoming (outgoing) muon and
internal neutrino with b (s) and t quarks in the two-loop diagrams for $%
a_{\mu}$, respectively, the decay of $b\rightarrow s \gamma$ can be
generated. For a rough estimation, the Wilson coefficient $C_7^{2-loop}$
associated with the operator $\bar{s}i\sigma_{\mu\nu}P_{R}b$ is positive,
which has an opposite sign to that of the SM. That is, two-loop effects on $%
b\rightarrow s \gamma$ can reduce the other possible new physics
contribution which are constructive with the SM so that $B\rightarrow X_s
\gamma$ could still be consistent with the experimental data in the model.
The detail analysis including one and two loops for $B\rightarrow X_s \gamma$
is beyond the scope of this paper.

\begin{figure}[tbp]
\hspace{1.5cm} \psfig{figure=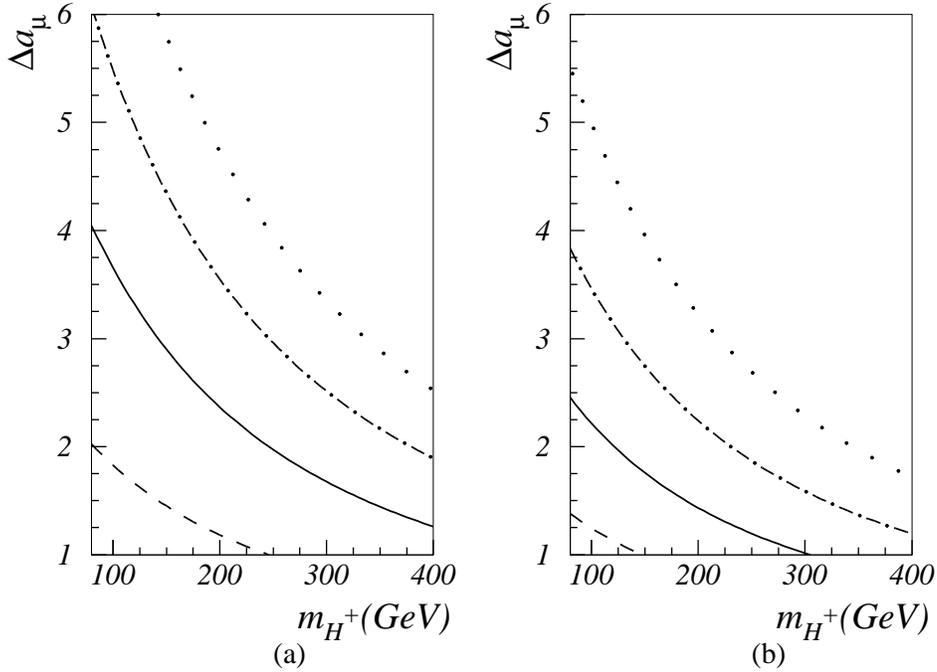,height=3.5in }
\caption{$\Delta{a}_{\mu}$ (in units of $10^{-9}$) as a function
of the charged Higgs mass with $\mu=2\;TeV$ and
$m_{\tilde{q}_{1}}=110 \ GeV$. The dashed, solid, dot-dashed and
dotted lines stand for (a) $\hat{A}_{b}=-1.5$,
$-3.0$, $-4.5$, $-6.0$ with $\tan\beta=40$ and (b) $\tan \beta=30$, $40$, $%
50 $ and $60$ with $\hat{A}_{b}=-2.0$, respectively. }
\label{g2tan}
\end{figure}

\begin{figure}[tbp]
\hspace{1.5cm} \psfig{figure=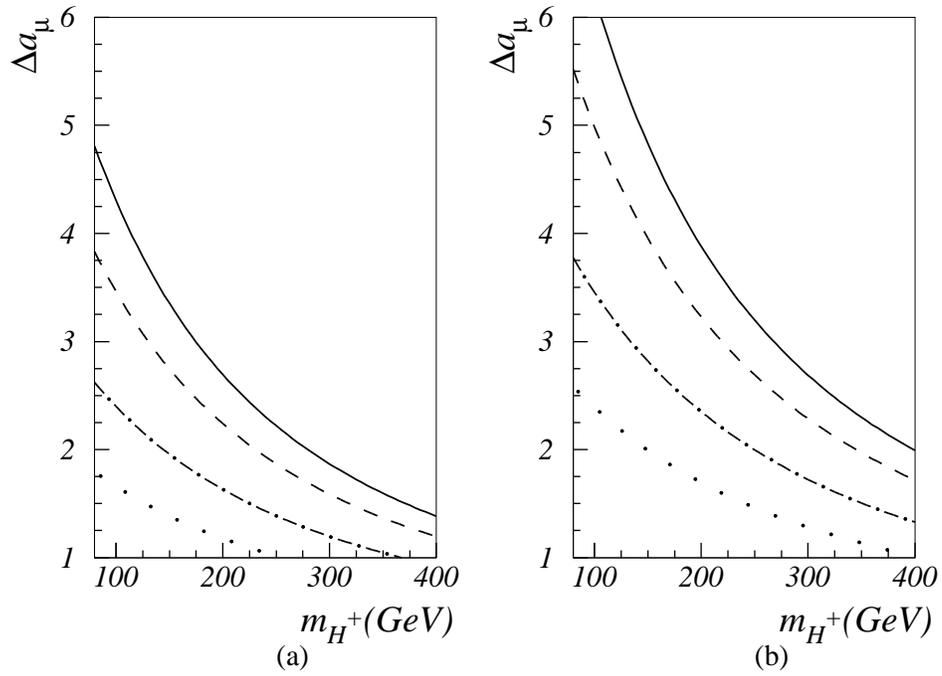,height=3.5in }
\caption{$\Delta{a}_{\mu}$ (in units of $10^{-9}$) as a function
of the charged Higss mass with $\mu=2\ TeV$ and
$\hat{A}_{b}=-2.0$. The dashed, solid, dot-dashed and dotted lines
stand for $m_{\tilde{q}_{1}}=90$, $110$, $150$ and $200\ GeV$,
with (a) $\tan\beta=50$ and (b) $\tan \beta=60$, respectively. }
\label{g2mst}
\end{figure}

In Figures \ref{g2tan} and \ref{g2mst}, we show how the SUSY parameters
enter in the contributions to $\Delta a_{\mu}$. The results can be
summarized as follows:

\begin{enumerate}
\item  In Eq. (\ref{g-2}), we factor out the $\mu $ parameter as the
definition of a proper scale for the low energy SUSY so that the ratio $%
\hat{A}_{b}$ could be the guideline of the different scale needed between
the electroweak and SUSY breaking. From Figure \ref{g2tan}(a), in terms of
the lower bound of $2.6\sigma $ level for $\Delta a_{\mu }$, we know that $%
\mu $ can be $2\ (0.5)\ TeV$ while $|A_{b}|_{min}\sim 9.0\ (36)\ TeV$ for $%
\tan \beta =40$ and $M_{H^{+}}=400\ GeV$. We note that if we use a larger $%
\tan \beta $ and lighter $M_{H^{+}}$, the SUSY soft breaking parameter, $%
A_{b}$, can be further reduced. Following the analysis in Refs. \cite{DGG}
and \cite{CGNW}, we choose sign($\mu $)$>$0 and sign($A_{b}$)=sign ($A_{t}$)$%
<$0 to satisfy the bound of $B\rightarrow X_{s}\gamma $.

\item  From Figure \ref{g2tan}(b), it is clear that $\Delta a_{\mu }$
strongly depends on the value of $\tan \beta $. This is because that Eq. (%
\ref{g-2}) is associated with a squared $\tan \beta $ arising from both
couplings of $H^{+}\tilde{t}_{L}^{*}\tilde{b}_{R}$ and $H^{+}\bar{\nu}%
_{L}\ell _{R}$. This leads to the contribution increased by a factor 2 when
replacing $\tan \beta =40$ by 60.

\item  According to Figure \ref{g2mst}, if we use $\tan \beta \sim O(50)$
and the bound in Eq. (\ref{newphys'}), the squark mass can be as heavy as $%
150\ GeV$, while the charged Higgs mass is fixed to be $200\ GeV$. As known,
the bound can be relaxed if the allowed SUSY breaking scale is higher.
\end{enumerate}

Finally, we remark that the neutral Higgs can also contribute to $\Delta
a_{\mu }$ through the coupling of the neutral Higgs and squarks similar to
the charged Higgs mechanism above, given by
\begin{eqnarray}
{\cal L}_{\tilde{f}\tilde{f}H^{0}} &=&\frac{g}{2M_{W}\sin \beta }%
(m_{t}A_{t}\sin \alpha +\mu \cos \alpha ))(\tilde{t}_{L}^{*}\tilde{t}_{R}+%
\tilde{t}_{R}^{*}\tilde{t}_{L})H^{0}  \nonumber \\
&+&\frac{g}{2M_{W}\cos \beta }(m_{b}A_{b}\cos \alpha +\mu \sin \alpha )(%
\tilde{b}_{L}^{*}\tilde{b}_{R}+\tilde{b}_{R}^{*}\tilde{b}_{L})H^{0}.
\label{nhiggs}
\end{eqnarray}
In this case, there is no suppression arising from the W-boson mass unlike
that with the charged Higgs. Therefore, the neutral Higgs effect usually can
be larger than the charged Higgs one with the assumption of the same masses.
The results can be easily obtained by setting $M_{W}=0$ and substituting the
relevant couplings in Eq. (\ref{g-2}). Following our analysis above, we
expect that the neutral scalar mass can be as heavy as $100\ GeV$ in
contrast with the case of the non-SUSY two Higgs doublet model (model II)
where a light scalar mass, $M_{H^{0}}\le 5$ GeV, is inevitable. For a
pseudoscalar boson, due to the opposite sign in the couplings of the
different chiral squarks, given by $(\tilde{q}_{L}^{*}\tilde{q}_{R}-\tilde{q}%
_{R}^{*}\tilde{q}_{L})A^{0}$, the contribution to $a_{\mu }$ vanishes. On
the contrary, if CP violating source is from the $\mu $ and $A_{t,b}$ terms,
the CP violating observables, such as electric dipole moments (EDMs) of
fermions, can arise from diagrams with the pseudoscalar.

In sum, we have analyzed the contribution of a generic charged Higgs to the
muon anomalous magnetic moment in the SUSY model. We have illustrated that
the experimental value of $a_{\mu }$ can be explained by the two-loop
charged Higgs diagrams without a further fine tuning and the allowed
parameter spaces are relatively large. For evading the strong constraints of
$B\rightarrow X_{s}\gamma $ on $m_{H^{+}}$, the chargino and squarks are as
light as charged Higgs and these conditions are detectable in present and
future colliders. Due to the enhancements of $A_{b}$ and $\tan \beta $, the
mass of the charged Higgs could be over $400\ GeV$ with proper values of
other parameters.\\

\noindent {\bf Acknowledgments}

We would like to thank We-Fu Chang for useful discusses. This work was
supported in part by the National Science Council of the Republic of China
under Contract Nos. NSC-89-2112-M-007-054 and NSC-89-2112-M-006-033 and the
National Center for Theoretical Science.

\newpage
\begin{figcap}
\item
Feynman diagrams for effective vertices of
$H^{+}-\gamma-W^{+}$ where squarks are in the internal loops.

\item
$\Delta{a}_{\mu}$ (in units of $10^{-9}$) as a function of the
charged Higgs mass with $\mu=2\;TeV$ and $m_{\tilde{q}_{1}}=110 \
GeV$. The dashed, solid, dot-dashed and dotted lines stand for
 (a)
 $\hat{A}_{b}=-1.5$, $-3.0$, $-4.5$, $-6.0$ with $\tan\beta=40$ and (b)
$\tan \beta=30$, $40$, $50$ and $60$ with $\hat{A}_{b}=-2.0$,
respectively.

\item
$\Delta{a}_{\mu}$ (in units of $10^{-9}$) as a function of the
charged Higss mass with $\mu=2\ TeV$ and $\hat{A}_{b}=-2.0$. The
dashed, solid, dot-dashed and dotted lines stand for
$m_{\tilde{q}_{1}}=90$, $110$, $150$ and $200\ GeV$, with (a)
$\tan\beta=50$ and (b) $\tan \beta=60$, respectively.
\end{figcap}

\newpage

\end{document}